\documentclass{aa} 
\usepackage{graphicx} 
\begin{document} 
\newcommand{\lsim}{\stackrel{\scriptstyle <}{\phantom{}_{\sim}}} 
\newcommand{\gsim}{\stackrel{\scriptstyle >}{\phantom{}_{\sim}}} 
%   \thesaurus{08     % A&A Section 8: Stars 
%              (02.04.1;  % Dense matter, 
%               08.05.3;  %  
%               08.09.3;  % Stars: interiors,     
%               08.14.1)} % Stars: neutron.       
% 
   \title{Cooling of Hybrid Neutron Stars and Hypothetical  
        Self-bound Objects with Superconducting Quark Cores} 
 
%   \subtitle{} 
 
   \author{D. Blaschke \inst{1,2} 
        \and H. Grigorian \inst{1,3} 
%\thanks{Research supported in part by the Volkswagen Stiftung 
%under 
%        grant no.\ I/71 226 and by DFG under grant no. 436 ARM 17/1/00} 
%        \and T. Kl\"ahn \inst{1} 
%       \and A. Sedrakian \inst{3}  
%       \fnmsep\thanks{Just to show the usage 
%          of the elements in the author field} 
        \and D.N. Voskresensky \inst{4}} 
 
   \offprints{D. Blaschke} 
 
   \institute{Fachbereich Physik, Universit\"at Rostock, 
        Universit\"atsplatz 1, D--18051 Rostock, Germany\\ 
         email: blaschke@darss.mpg.uni-rostock.de  
	\and 
	European Centre for Theoretical Studies ECT$^*$, Villa Tambosi,  
	Strada delle Tabarelle 286, 38050 Villazzano (Trento), Italy  
        \and  
	Department of Physics, Yerevan State University, Alex 
        Manoogian Str. 1, 375025 Yerevan, Armenia\\ 
        email: hovik@darss.mpg.uni-rostock.de 
        \and  
	Moscow Institute for Physics and Engineering,  
	Kashirskoe shosse 31, 115409 Moscow, Russia\\ 
        Gesellschaft f\"ur Schwerionenforschung GSI, 
         Planckstrasse 1, D-64291 Darmstadt, Germany\\ 
        email: D.Voskresensky@gsi.de 
             } 
 
   \date{Received 14 September 2000 / Accepted 02 January 2001} 
	 
   \titlerunning{Cooling of Hybrid Neutron Stars ...} 
   \authorrunning{Blaschke, Grigorian \& Voskresensky} 
  
   \abstract{ 
We study the consequences of superconducting quark cores (with  
color-flavor-locked phase as representative example) for evolution of  
temperature profiles and the cooling curves in quark-hadron  hybrid stars  
and in hypothetical self-bounded objects having no a hadron shell (quark core  
neutron stars).  
The quark gaps are varied from $0$ to $\Delta_q =50$ MeV. 
For hybrid stars we find time scales of $1\div5$, $5\div10$ and $50\div100$  
years for the formation of a quasistationary temperature distribution in the  
cases $\Delta_q =0$, 0.1  MeV and $\gsim$ 1 MeV, respectively. 
These time scales are governed by the heat transport within quark cores   
for large diquark gaps ($\Delta \gsim$ 1 MeV) and within the hadron shell for  
small diquark gaps ($\Delta \lsim 0.1$ MeV).  
For quark core neutron stars we find a time scale $\simeq 300$ years for the  
formation of a quasistationary temperature distribution in the case 
$\Delta \gsim$ 10 MeV and a very short one for $\Delta \lsim$ 1 MeV.  
If hot young compact objects will be observed they can be interpreted as  
manifestation of large gap color superconductivity. 
Depending on the size of the pairing gaps, the compact star takes different  
paths in the $\mbox{lg}(T_s) $ vs. $\mbox{lg}(t)$ diagram where $T_s$ is the  
surface temperature.  
Compared to the corresponding hadronic model which well fits existing data 
the model for the hybrid neutron star (with a large diquark gap) shows too  
fast cooling.  
The same conclusion can be drawn for the corresponding self-bound objects. 
 } 
   \maketitle 
%________________________________________________________________ 
 
\section{Introduction} \label{sec:intro} 
 
The interiors of compact stars are considered as systems where  
high-density phases of strongly interacting matter do occur in nature,  
see Glendenning (1996) and Weber (1999) for recent textbooks. 
The consequences of different phase transition scenarios for the cooling  
behaviour of compact stars have been reviewed recently  
in comparison with existing X-ray data, see Page (1992), Schaab et al. (1997). 
 
A completely new situation might arise if the scenarios suggested for 
(color) superconductivity (\cite{arw98,r+98}) with large diquark pairing gaps  
($\Delta_q \sim 50 \div 100$ MeV) in quark matter are applicable to neutron  
star interiors.  
Various phases are possible. The two-flavor (2SC) or the three-flavor (3SC)  
superconducting phases allow for unpaired quarks of one color whereas in 
the 
color-flavor locking (CFL) phase all the quarks are paired. 
 
Estimates of the cooling evolution have been performed (\cite{bkv99})  
for a self-bound isothermal 
quark core neutron star (QCNS) which has a crust but no hadron shell,  
and for a quark star (QS) which has neither crust nor hadron shell.  
It has been shown there in the case of the 2SC (3SC) phase of QCNS 
that the consequences of the occurrence of gaps for the cooling curves are  
similar to the case of usual hadronic neutron stars (enhanced cooling).  
However, for the CFL case it has been shown that the cooling is 
extremely fast since the drop in the specific heat of  
superconducting quark matter dominates over the reduction of the  
neutrino emissivity.  
As has been pointed out there, the abnormal rate of 
the temperature drop is the consequence of the approximation of homogeneous  
temperature profiles the applicability of which should be limited by the heat  
transport effects. 
Page et al. (2000) estimated the cooling of hybrid neutron stars (HNS) where  
heat transport effects within the superconducting quark core have been  
disregarded.  
Neutrino mean free path in color superconducting quark matter have been  
discussed in (\cite{cr00}) where a short period of cooling delay at the onset  
of color superconductivity for a QS has been conjectured in accordance with 
the estimates of (\cite{bkv99}) in the CFL case for small gaps.  
 
In the present paper we want to consider a more detailed  
scenario for the thermal evolution of HNS and QCNS 
which includes the heat transport in both the quark and the hadronic matter. 
We will demonstrate how long it takes for the  HNS and for the QCNS to  
establish a quasistationary temperature profile and then we consider the  
influence of both the diquark pairing gaps and the hadronic gaps 
on the evolution of the surface temperature.  
 
\section{Processes in HNS and QCNS} 
\subsection{Processes in quark matter} 
\subsubsection{Emissivity} 
A detailed discussion of the neutrino emissivity of quark matter 
without the possibility of  color superconductivity has  been given  
first by Iwamoto (1982).  
In his work the quark direct Urca (QDU) 
reactions $d\rightarrow ue\bar{\nu}$ and $ue\rightarrow d {\nu}$ 
have been suggested as the most efficient processes.  
Their emissivities have been obtained as 
\footnote{Please notice that the numerical factors in all the estimates 
below depend essentially on rather unknown factors 
and thereby can be still varied.}  
\begin{eqnarray}\label{neutr-DU} 
\epsilon^{\rm QDU}_\nu \simeq 9.4 \times 
10^{26}\alpha_s u Y_e^{1/3}\zeta_{\rm QDU}~ 
T_{9}^6~ {\rm erg~cm^{-3}~s^{-1}}, 
\end{eqnarray} 
where at a compression $u=n/n_0\simeq 2$ the  
strong coupling constant is $\alpha_s \approx 1$ and decreases  
logarithmically at still higher densities. 
The nuclear saturation density is $n_0=0.17~ {\rm fm}^{-3}$,  
$Y_e =n_e /n$ is the electron fraction, and 
$T_9$ is the temperature in units of $10^9$ K.  
If for somewhat larger density the electron fraction was too small  
($Y_e<Y_{ec}\simeq 10^{-8}$), 
then all the QDU processes  would be completely 
switched off (\cite{DSW83}) and the neutrino emission would be 
governed by two-quark reactions like the quark modified Urca (QMU) 
and the quark bremsstrahlung (QB) processes 
$dq\rightarrow uqe\bar{\nu}$ and $q_1 q_2 \rightarrow q_1 q_2 \nu\bar{\nu}$, 
respectively.  
The emissivities of the QMU and QB processes have been estimated as 
(\cite{I82}) 
\begin{eqnarray}\label{neutr-B} 
\epsilon^{\rm QMU}_\nu\sim \epsilon^{\rm QB}_\nu  \simeq 9.0\times  
10^{19}\zeta_{\rm QMU}~T_{9}^8~ {\rm erg~cm^{-3}~s^{-1}}. 
\end{eqnarray} 
Due to the pairing, the emissivities of QDU processes 
are suppressed by a factor $\zeta_{\rm QDU} \sim \mbox{exp}(-\Delta_q /T)$ 
and the emissivities of QMU and QB processes are suppressed by a  
factor $\zeta_{\rm QMU} \sim \mbox{exp}(-2\Delta_q /T)$ for  
$T<T_{{\rm crit},q}\simeq 0.4~\Delta_q$ whereas for  
$T>T_{{\rm crit},q}$ these factors are equal to unity.  
The modification of $T_{{\rm crit},q}(\Delta_q )$ relative to the standard  
BCS formula is due to the formation of correlations as, e.g., instanton- 
anti-instanton molecules (\cite{r+99}).  
For the temperature dependence of the gap below $T_{{\rm crit},q}$ we use the  
interpolation formula $\Delta(T)=\Delta(0) \sqrt{1-T/T_{{\rm crit},q}}$, with  
$\Delta(0)$ being the gap at zero temperature. 
 
The contribution of the reaction $ee\rightarrow ee\nu\bar{\nu}$ 
is very small (\cite{KH99}) 
\begin{equation} 
\epsilon^{ee}_\nu = 2.8\times 10^{12}\, Y_e^{1/3} u^{1/3} T_9^8~ 
{\rm erg~cm^{-3}~s^{-1}},   
\label{eq:4} 
\end{equation} 
but it can become important when quark processes are blocked out for large  
values of $\Delta_q/T$ in superconducting quark matter. 
 
\subsubsection{Specific heat}  
 
For the quark specific heat we use 
an expression (\cite{I82}) 
\begin{eqnarray}\label{heat} 
c_{q}\simeq 10^{21}u^{2/3}\zeta_{\rm S}~T_9~{\rm erg~cm^{-3}~K^{-1}},  
\end{eqnarray} 
where $\zeta_{\rm S}\simeq 3.1~(T_{{\rm crit},q}/T)^{5/2}~\mbox{exp} 
(-\Delta_q /T)$. 
Besides, one should add the gluon-photon contribution (\cite{bkv99}) 
\begin{eqnarray}\label{gluon} 
c_{g-\gamma}= 3.0\times 10^{13}~N_{g-\gamma}~T_9^3~{\rm erg~cm^{-3}~K^{-1}}, 
\end{eqnarray} 
where $N_{g-\gamma}$ is the number of available  massless gluon-photon states 
(which are present even in the color superconducting phase),   
as well as the electron one, 
\begin{equation}\label{e} 
c_e= 5.7\times 10^{19} \,Y_e^{2/3} u^{2/3}~T_9~{\rm erg~cm^{-3}~K^{-1}}. 
\end{equation} 
 
\subsubsection{Heat conductivity}  
 
The heat conductivity of the matter is the sum of 
partial contributions (\cite{flowers}) 
\begin{equation} 
 \label{conductivity2} 
\kappa = \sum_{i}\kappa_{i},\,\,\, 
\frac{1}{\kappa_i} = \sum_{j}\frac{1}{\kappa_{ij}}~, 
\end{equation} 
where $i,j$ denote the components (particle species). 
For quark matter $\kappa$ is the sum of the partial conductivities of the 
electron, quark and gluon components (\cite{HJ89,BH99}) 
\begin{equation} 
 \label{total} 
\kappa = \kappa_{e} + \kappa_{q}+\kappa_{g}, 
\end{equation} 
where 
$\kappa_{e}\simeq \kappa_{ee}$ is determined by electron-electron scattering 
processes since in superconducting quark matter the partial contribution  
$1/\kappa_{eq}$ (as well as $1/\kappa_{gq}$ ) is additionally suppressed by a  
$\zeta_{\rm QDU}$ factor, as for the scattering on impurities in  
metallic superconductors.  
For  $\kappa_{ee}$ we have   
\begin{eqnarray}\label{ee} 
\kappa_{ee}&=&5.5\times 10^{23}u~{Y_e}~{T_9}^{-1}~ 
{\rm erg~s^{-1}cm^{-1}K^{-1}}, 
\end{eqnarray} 
and 
\begin{eqnarray} 
\kappa_q &\simeq&  \kappa_{qq}\nonumber\\ 
&\simeq& 
1.1\times 10^{23}\sqrt{\frac{4\pi}{\alpha_s}}~u~\zeta_{\rm QDU}{T_9}^{-1} 
{\rm erg~s^{-1}cm^{-1}K^{-1}}, 
\end{eqnarray} 
where we have accounted for the suppression factor.   
The contribution of massless gluons we estimate as 
\begin{eqnarray} 
\kappa_{g}&\simeq&\kappa_{gg}\nonumber\\ 
&\simeq& 6.0 \times 10^{17}~T_9^{2}~{\rm erg~s^{-1}cm^{-1}K^{-1}}.  
\end{eqnarray} 
 
\subsection{Processes in hadronic matter} 
 
\subsubsection{Emissivity} 
 
DU processes $n\rightarrow pe\bar{\nu}$, $pe\rightarrow n\nu$ 
can occur when the proton fraction exceeds 11\% (\cite{LPPH91}). 
It does not occur for the equation of state we use but there are examples  
for the opposite case (\cite{p+00}). This is another difference of our 
present work to the recent one of Page et al. (2000). 
 
Next, we take into account the modified Urca (MU) processes  
$nn\rightarrow npe\bar{\nu}$, $np\rightarrow ppe\bar{\nu}$, and the reverse  
processes.  
The final emissivity is given by (\cite{FM79,YLS99}) 
\begin{eqnarray} 
\label{eq:1} 
\epsilon_{\nu}^{\rm nMU}&=&8.6\times 10^{21}m^{*4}_{\rm nMU}   
(Y_e u)^{1/3}\zeta_{\rm nMU}T_9^8 {\rm erg cm^{-3}s^{-1}},\\ 
\epsilon_{\nu}^{\rm pMU}&=&8.5\times 10^{21}m^{*4}_{\rm pMU} 
(Y_e u)^{1/3}\zeta_{\rm pMU}T_9^8 {\rm erg cm^{-3}s^{-1}}. 
\end{eqnarray} 
Here $m_i^* =\sqrt{m_{{\rm{rel}},i}^{*2}+p_{{\rm{F}},i}^2}$ 
is the non-relativistic quasiparticle effective mass related to the  
in-medium one-particle energies from a given 
relativistic mean field model for $i=n,p$.  
We have introduced the abbreviations  
$m^{*4}_{\rm nMU}=({m_n^*}/{m_n})^3({m_p^*}/{m_p})$ and  
$m^{*4}_{pMU}=({m_p^*}/{m_p})^3({m_n^*}/{m_n})$.  
The suppression factors are 
$\zeta_{\rm nMU}=\zeta_n \zeta_p 
\simeq {\exp}\{-[\Delta_n(T)+\Delta_p(T)]/T\}$, 
$\zeta_{\rm pMU}\simeq \zeta_p^2$, 
and should be replaced by unity for $T>T_{{\rm crit},i}$ when for given  
species $i$ the corresponding gap vanishes. 
For neutron and proton $S$-wave pairing,  
$\Delta_{i}(0)=1.76~T_{{\rm crit},i}$   
and for $P$-wave pairing of neutrons $\Delta_n(0) =1.19~ T_{{\rm crit},n}$. 
To be conservative we have used in (\ref{eq:1}) the free one-pion exchange  
estimate of the $NN$ interaction amplitude.  
Restricting ourselves to a qualitative analysis we use here simplified 
exponential suppression factors $\zeta_i$. 
In a more detailed analysis these $\zeta_i$-factors have prefactors with 
rather strong temperature dependences \cite{YLS99}. At temperatures $T\sim T_c$
their inclusion only slightly affects the resulting cooling curves.
For $T\ll T_c$ the MU process gives in any case a negligible contribution to 
the total emissivity and thereby corresponding modifications can again be 
omitted. Also for the sake of simplicity the general possibility of a 
$^3$P$_2(|m_J|=2)$ pairing which may result in a power-law behaviour of the 
specific heat and the emissivity of the MU process \cite{VS87,YLS99} is 
disregarded since mechanisms of this type of pairing are up to now not 
elaborated.
Even more essential modifications of the MU rates may presumably come 
from in-medium effects which could result in extra prefactors of 
$10^2 \div 10^3$ already at $T \sim T_c$.

In order to estimate the role of the in-medium effects in the $NN$  
interaction for the HNS cooling we have also performed calculations for the  
so-called medium modified Urca (MMU) process (\cite{VS86,MSTV90}) by 
multiplying the rates (\ref{eq:1}) by the appropriate prefactor  
\begin{equation}\label{VS-FM} 
\epsilon_{\nu}^{\rm MMU}/\epsilon_{\nu}^{\rm MU}\simeq 10^{3} 
\left[ 
\Gamma^{6}(g^{\prime})/\widetilde{\omega}^{8}(k\simeq p_F )\right] 
u^{10/3}, 
\end{equation} 
where the value $\Gamma (g^{\prime}) 
\simeq 1/[1+1.4~u^{1/3}]$ is due to the dressing of $\pi NN$ 
vertices 
and $\widetilde{\omega}\leq m_\pi$  
is the effective pion gap which we took as function of density 
from Fig. 2 of (\cite{sch+97}). 
 
For $T<T_{\rm crit}$ the most important contribution comes from the 
neutron (\cite{FRS76,VS87,YLS99}) and the proton (\cite{VS87}) pair 
breaking and formation processes. 
Their emissivities we take from Ref. (\cite{sch+97}) which is applicable for 
both the cases of $S$- and $P$-wave nucleon pairing 
\begin{eqnarray} 
\label{eq:2} 
\epsilon_{\nu}^{\rm NPBF}&=&6.6\times 10^{28} ({m_n^*}/{m_n}) 
({\Delta_n(T)}/{\rm MeV})^7~u^{1/3}\nonumber \\ 
&&\times \xi~I({\Delta_n(T)}/{T})~{\rm erg~ cm^{-3}s^{-1}},\\ 
\epsilon_{\nu}^{\rm PPBF}&=&0.8\times 10^{28} ({m_p^*}/{m_p}) 
({\Delta_p(T)}/{\rm MeV})^7~u^{2/3}\nonumber \\ 
&&\times ~I({\Delta_p(T)}/{T})~{\rm erg~cm^{-3}s^{-1}}, 
\end{eqnarray} 
where 
\begin{equation} 
I({\Delta_i(T)}/{T})\simeq 0.89~\sqrt{T/\Delta_i(T)} 
\exp[-2 \Delta_i(T)/T]~, 
\label{eq:3} 
\end{equation} 
$\xi \simeq 0.5$ for $^1S_0$ pairing 
and $\xi \simeq 1$ for $^3P_2$ pairing. 
%\begin{equation} 
%\xi=\left\{ 
%\begin{array}{ll} 
%0.45& \quad 1S_0~,\\ 
%1& \quad 3P_2~. 
%\end{array} 
%\right. 
%\end{equation} 
A significant contribution of the proton channel is due to the $NN$  
correlation effect taken into account in (\cite{VS87}). 
 
\subsubsection{Specific heat} 
 
For the nucleons ($i=n,p$), the specific heat is (\cite{M79}) 
\begin{equation} 
c_i= 1.6\times 10^{20}({m_i^*}/{m_i})~u^{1/3} \zeta_{\rm MMU}~T_9~ 
{\rm erg~cm^{-3}K^{-1}}. 
\end{equation} 
The specific heat for the electrons is determined by (\ref{e}) and that for  
the photons for $T>T_{cp}$ is given by (\ref{gluon}) with the corresponding  
number of polarizations $N_\gamma =2$. 
 
\subsubsection{Heat conductivity} 
 
The total conductivity is 
\begin{equation} 
 \label{total_n} 
\kappa \simeq \kappa_{n}+\kappa_{e},\,\,\, 
\frac{1}{\kappa_n} = \frac{1}{\kappa_{nn}} + \frac{1}{\kappa_{np}}, 
\end{equation} 
where the partial contributions are (\cite{flowers,BH99}) 
\begin{eqnarray} 
\kappa_{nn}&=&8.3\times 10^{22} \left(\frac{m_n}{m_n^*}\right)^4 
 \frac{z_n^3~\zeta_n}{S_{kn}~T_9}~ 
{\rm erg~ s^{-1}cm^{-1}K^{-1}}, \\[8pt] 
S_{kn}&=&0.38~z_n^{-7/2} + 3.7~z_n^{2/5}~,  
\\ 
\kappa_{np}&=&8.9\times 10^{16} \left(\frac{m_n}{m_n^*}\right)^2  
\frac{z_n^2\zeta_{p}T_9}{z_p^{3}S_{kp}}~{\rm erg~s^{-1}cm^{-1}K^{-1}},\\ 
S_{kp}&=&1.83~z_n^{-2} + 1.43~z_n^2(0.4+z_n^8)^{-1}~. 
\end{eqnarray} 
Here we have introduced the apropriate suppression factors $\zeta_i$ which act 
in the presence of gaps for superfluid hadronic matter and we have used the 
abbreviation $z_i=(n_i/(4 n_0))^{1/3}$. 
The heat conductivity of electrons is given by Eq. (\ref{ee}). 
 
\subsection{Processes in the mixed phase} 
 
In some density interval $n_{\rm H}<n<n_{\rm Q}$ we have a mixed phase in the 
HNS case. 
Although our treatment of the contribution of the mixed phase to the equation  
of state is rather poor, for the discussion of the neutrino processes  
we assume the mixed phase to be constructed as a lattice (for $T<T_{\rm melt}$) 
of color superconducting quark droplets embedded in the nucleonic Fermi sea 
(superfluid for $T<T_{{\rm crit},i}; i=n,p$) at densities near the 
lower boundary density $n_{\rm H}$ which is reorganized as a  
lattice of nucleon droplets in the quark 
Fermi sea near the upper boundary density $n_{\rm Q}$.  
We suppose that the 
mixed phase contributes to the neutrino reactions and to the heat transport 
as two homogeneous phases weighted according to their partial volumina and  
thus suppress reactions on the impurities which could be as 
efficient as one-nucleon/quark processes in the liquid phase (\cite{RBP00}).  
Thus we assume $T<T_{\rm melt}$ in our case.  
 
\subsection{Crust and surface} 
 
In order to avoid a complicated analysis of the processes in the crust 
we use an interpolated relation between the surface and the  
crust temperatures (see \cite{ST83})  
$T_{s} =(10~T_m )^{2/3}$, 
where temperatures are given in K, 
the so-called Tsuruta law \cite{T79}.
More elaborated and complicated dependences of $T_s$ on $T_m$ and the 
parameters of the crust can be found in (\cite{GPE82,GPE83,V88,PCY97}).
As we have estimated, their inclusion does not affect qualitatively the 
conclusions of this work.
The mantle temperature $T_{m}=T(r_m)$ we take at the distance $r_m$ from the  
center where the density is equal to the critical one for the transition 
to the mixed (Aen) phase.  
This is correct for time scales exceeding those determined  by the heat  
conductivity of the crust estimated as $\lsim 10$ yr for hadronic stars and  
$\lsim 1$ yr for strange stars (see \cite{Pizzochero 1991,LPPH91}).  
Since the crust is rather thin and has a much smaller heat content than the 
bulk of the star its contribution to the cooling delay for time intervals  
shorter than the ones given above is negligible. 
Due to these reasons and for the sake of 
simplicity we disregard these effects in further considerations.  
 
We also add  the photon contribution to the emissivity at the surface 
\begin{eqnarray}\label{photon} 
\epsilon_{\gamma} \simeq 2\times 10^{18}\left(\frac{R}{10~{\rm 
km}}\right)^{-1}~ 
T_{s7}^4~{\rm erg}~{\rm cm}^{-3}~{\rm sec}^{-1},  
\end{eqnarray}  
where $T_{s7}$ is the surface temperature in units of $10^7$ K, 
%see \cite{ST83} 
using it as the boundary condition in our transport code.  
Thus simplifying we assume that the outer crust 
and the photosphere are thin enough to approximate $r_m\simeq R$, where 
%but the law $T_s=(10 T_m)^{2/3}$ \cite{ST83,T79} is the model employed 
%for the  
%thermal properties of the outer crust, where  
$R$ is the star radius. 
%% of the star and $T_s$ the surface temperature. 
Then, the luminosity at $r_m$ corresponds to the surface luminosity 
of the photons from the star $l(r_m)=L_\gamma(T_s)$.

%%%%%%%%%%%%%%%%%%%%%%%%%%%%%%%%%%%%%%%%%%%%%%%%%%%%%%%%%%%%%%%%%%%%%%%%% 
\section{Evolution of inhomogeneous emissivity profiles} 
%%%%%%%%%%%%%%%%%%%%%%%%%%%%%%%%%%%%%%%%%%%%%%%%%%%%%%%%%%%%%%%%%%%%%%%%% 
 
\subsection{Equation of state and gaps} 
 
The equation of state is decribed according to the model (\cite{cgpb}) 
incorporating three possible phases (depending on the star mass, the 
value of the bag constant $B$ and other parameters as the mass of the strange 
quark, the compressibility of hadron matter etc.).  
These phases are hadronic matter, quark-hadron mixed phase and pure quark  
matter. 
As a representative example we consider below a star with a mass of   
1.4 $M_{\odot}$ for the neutron star consisting of pure hadronic matter  
(radius $\simeq 13 $ km), for the HNS (radius $\simeq 10 $ km) and for the  
self-bound QCNS (radius $\simeq 9$ km). 
The typical electron fraction is $Y_e\simeq 0.04 $ for the hadronic star which 
does not allow for the occurrence of direct Urca processes. 
  
For the HNS the electron fraction is 
%$Y_e\simeq 0.01 $ for $n=n_0$,  
$Y_e \simeq 0.03 $ for $n=n_{\rm H}$, where $n_{\rm H}$ corresponds to the  
critical density when first seeds of quark phase appear;  
$Y_e \simeq 8.5 \times 10^{-5} $ for $n=n_{\rm Q}$ when hadron seeds  
disappear, i.e. at the boundary of the quark core,   
and $Y_e \simeq 2 \times 10^{-5} $ for $n=n_{c}$, i.e. for the 
baryon density in the star center. 
%The DU processes are forbidden in both hadronic and quark phases.  
 
The radius corresponding to $n=n_{\rm H}$ is $r_{\rm H}=8$ km and that  
corresponding to $n=n_{\rm Q}$ is $r_{\rm Q}=6.5$ km.  
We take the same dependences $\Delta_{i}(n)$ for $i=n,p$ as in  Fig. 6 
of (\cite{sch+97}).  
As an alternative example we also consider the model with suppressed gaps 
to estimate the effect of possible uncertainties. 
In the mixed phase we take $\Delta_i =\Delta_{i}(n_{\rm H})$ 
for the hadron sub-system and $\Delta=\Delta_q $ for the quark one, 
and in the quark phase we for simplicity take $\Delta_q ={\rm const}$. 
  
For QCNS besides the case when $Y_e$ is determined from  our calculation 
we consider also $Y_e =0$ as an extreme example of the case  
$Y_e <Y_{ec}$, in order to understand its possible consequencies. Then 
direct Urca does not occur also in the quark phase. 
 
\subsection{Cooling} 
 
Due to the inhomogeneous distribution of matter  
inside the star and finite heat conductivity the temperature profile  
$T(r,t)$ during the cooling process can differ from the isothermal  
equilibrium one for which the temperature on the inner crust boundary  
($T_{m}$) and the central one ($T_c$) are connected by the relation  
$T_{m}=T_{c} \exp [\Phi(0) - \Phi(R)]$, 
where $\Phi(0)-\Phi(R)$ is the difference of the gravitational potentials  
in the center and at the surface of the star, respectively. 
 
The flux of energy $l(r)$ per unit time through a spherical slice at 
the distance $r$ from the center is proportional to the gradient of the  
temperature on both sides of the spherical slice, 
\begin{equation}\label{lr} 
l(r) = - 4 \pi r^2 \kappa (r) \frac{\partial (T{\rm e}^\Phi)}{\partial r}  
{\rm e}^{-\Phi} \sqrt{1-\frac{2 M}{r}}~, 
\end{equation} 
%where $\kappa=\kappa(r)$  is the coefficient of heat  
%conductivity and  
the factor  
${\rm e}^{-\Phi} \sqrt{1-\frac{2 M}{r}}$ 
corresponds to the relativistic correction of the time scale and the 
unit of thickness. 
The equations for energy balance and thermal energy transport are 
(\cite{W99}) 
\begin{eqnarray}\label{lbal} 
\frac{\partial }{\partial A}\left( l {\rm e}^{2 \Phi}\right)&=&  
- \frac{1}{n}\left( \epsilon_\nu {\rm e}^{2\Phi}  
+ c_V \frac{\partial }{\partial t} (T {\rm e}^\Phi)\right) ~,\\  
\frac{\partial }{\partial A}\left(T {\rm e}^{2 \Phi}\right)&=&  
- \frac{1}{\kappa} \frac{l {\rm e}^{\Phi}}{16 \pi ^2 r^4 n} ~, 
\label{Tbal} 
\end{eqnarray} 
where $n=n(r)$ is the baryon number density, $A=A(r)$ is the total baryon 
number in the sphere with radius $r$ and  
\begin{equation} 
\frac{\partial r}{\partial A}=\frac{1}{4 \pi r^2 n} \sqrt{1-\frac{2 
M}{r}}~. 
\end{equation} 
The total neutrino emissivity $\epsilon_\nu$ and the total specific heat 
$c_V$ are given as the sum of the corresponding partial contributions  
defined in the previous Section for a composition $n_i(r)$ of constituents  
$i$ of the matter under the conditions of the actual temperature profile 
$T(r,t)$.   
The accumulated mass $M=M(r)$ and the gravitational potential $\Phi=\Phi(r)$ 
can be determined by  
\begin{eqnarray}\label{potential} 
\frac{\partial M}{\partial A}&=&\frac{\varepsilon}{n} \sqrt{1-\frac{2 M}{r}}~, 
\\ 
\frac{\partial \Phi}{\partial A}&=&\frac{4 \pi r^3 p + m}{4 \pi r^2 n} 
\frac{1}{\sqrt{1-\frac{2 M}{r}}}~, 
\end{eqnarray} 
where $\varepsilon=\varepsilon(r)$ is the energy density profile 
and the pressure profile $p=p(r)$ is defined by the condition of 
hydrodynamical equilibrium  
\begin{equation}\label{tov} 
\frac{\partial p}{\partial A}=  
- (p + \varepsilon) \frac{\partial \Phi}{\partial A}~. 
\end{equation}  
The boundary conditions for the solution of (\ref{lbal}) and 
(\ref{Tbal}) read 
$l(r=0)=l(A=0)=0$ and $T(A(r_m)=A,t)=T(r_m=R,t)=T_m(t)$, respectively. 
 
In our examples we choose the initial temperature to be 1 MeV. 
This is a typical value for the temperature $T_{\rm opacity}$ 
at which the star becomes transparent for neutrinos. 
Simplifying we disregard the neutrino influence on transport.  
These effects dominate for $t \lsim 1 \div 100$ min, 
when the star cools down to $T \leq T_{\rm opacity}$ 
and become unimportant for later times. 
 
%%%%%%%%%%%%%%%%%%%%%%%%%%%%%%%%% Figure 1 %%%%%%%%%%%%%%%%%%%%%%%% 
\begin{figure} 
\resizebox{\hsize}{!}{\includegraphics{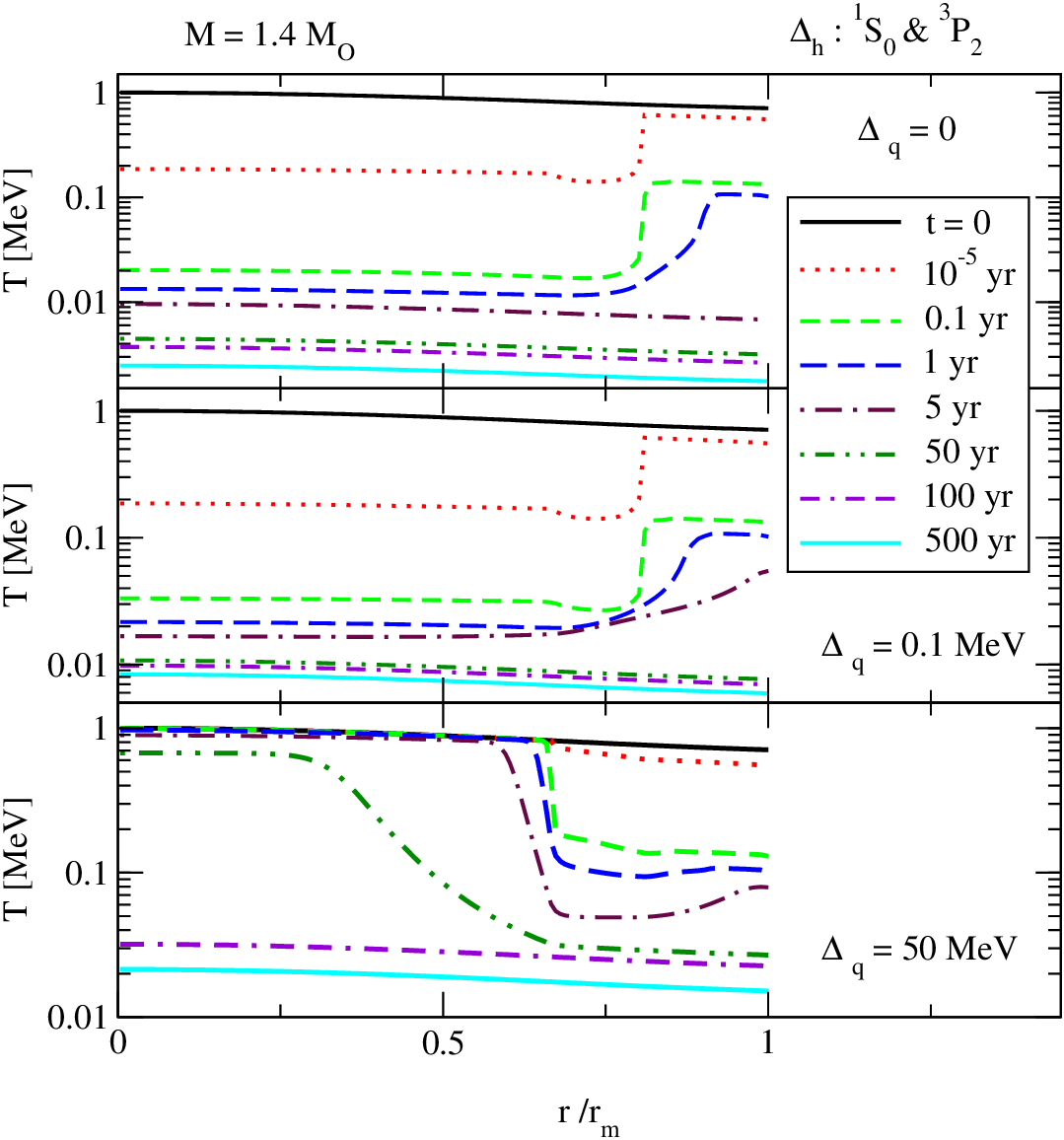}}
\caption{Early evolution of temperature profiles for a HNS of  
$M=1.4 M_\odot$  with large (lower panel), small (middle) and vanishing 
(upper panel) diquark pairing gaps.}  
\label{Tprof}
\end{figure} 
%%%%%%%%%%%%%%%%%%%%%%%%%%%%%%%%%%%%%%%%%%%%%%%%%%%%%%%%%%%%%%%%%% 
 
\section{Temperature evolution of HNS} 
 
\subsection{Temperature profiles $T(r,t)$} 
 
With the above inputs we have solved the evolution equation for the  
temperature profile.  
In order to demonstrate the influence of the size of the diquark and nucleon 
pairing gaps on the evolution of the temperature profile we have  
performed solutions with different values of the quark and nucleon gaps. 
The representative examples are shown in Fig. 1.  
Nucleon gaps are taken the same as in Fig. 6 of (\cite{sch+97}).  
The quark gaps larger than 1 MeV show the same typical behaviour as  
for the gap $\Delta_q =50$ MeV. The gaps much smaller than 1 MeV exhibit 
the typical small gap behaviour as for $\Delta_q =0.1$ MeV in our example.  
 
For large quark gaps, the hadron phase cools down 
on the time scale $t \simeq 0.1 \div 5$ yr due to the heat transport to the  
star surface whereas the quark core keeps the heat during this time. 
Therefore, if an independent measurement of the core  
temperature was possible, e.g. by neutrinos, then a core temperature stall  
during first several years of cooling  
evolution would be a case for quark core superconductivity with large 
pairing gaps. 
Then the quark core begins to cool down slowly from its mixed phase boundary,  
the cold from hadrons via mixed phase spreads to the center,  
demonstrating that the direct neutrino radiations during all the time 
are unefficient within the quark core. 
The homogeneous temperature profile is recovered at typical times of 
$t \simeq 50 \div 100$ yr.  
 
For $\Delta_q =0.1$ MeV the QDU processes and the heat conductivity within  
the quark core are quite efficient.  
Therefore at very small times ($t \ll 10^{-5}$ yr) the temperature profile  
within the quark core is a homogeneous one whereas during the next  
$5 \div 10$ yr the hadron shell is cooled down from the both sides, 
the mixed phase boundary at $n(r_{\rm H})=n_{\rm H}$ and the star surface. 
Thus, for $\Delta_q \ll 1$ MeV the interior 
evolution of the temperature has an influence on the surface temperature from  
the very early times. 
Actually the shortest time scale which determines the slowing of the 
heat transport within the crust is in this case $t \sim 1$ yr. 
Thus at much smaller times one has $T_s(t) \simeq T_s(0)$ with a  
drop of $T(r,t)$ in the vicinity of $r/r_m =1$. Due to the relatively small 
heat content of the crust this peculiar effect is disregarded here.  
 
Qualitatively the same behaviour is illustrated for normal quark matter 
with the only difference that quantitatively the quark processes are more 
efficient at $\Delta_q =0$ and the cold thereby traverses the hadron 
phase more rapidly, at typical times $t \simeq 1 \div 5$ yr.  
 
\subsection{Evolution of the surface temperature} 
 
A detailed comparison of the cooling evolution  
($\mbox{lg}~T_s$ vs. $\mbox{lg}~t$) of HNS for different values of 
quark and hadron gaps is given in Fig. 2. 
We have found that the curves for $\Delta_q \gsim 1$ MeV are very close to  
each other  demonstrating typical large gap behaviour.  
As representative example we again take $\Delta_q=50$ MeV. 
The behaviour of the cooling curve for $t \leq 50 \div 100$ yr is in straight 
correspondence with the heat transport processes discussed above.  
The subsequent time evolution is governed by the processes in the hadronic  
shell and by a delayed transport within the quark core with a dramatically  
suppressed neutrino emissivity from the color superconducting region.  
In order to demonstrate this feature we have performed a calculation with 
the nucleon gaps ($\Delta_i (n), ~i= n,p$) being artificially  suppressed  
by a factor 0.2. 
Then up to $\mbox{lg}(t[{\rm yr}]) \lsim 4$ the behaviour of the cooling  
curve is analogous to the one we would have obtained for pure hadronic matter. 
The curves labelled "MMU" are calculated with the rates of modified Urca  
processes of Eq. (\ref{eq:1}) multiplied by the factor (\ref{VS-FM}),  
i.e. with inclusion of appropriate medium modifications in the $NN$  
interaction.  
%%%%%%%%%%%%%%%%%%%%%%%%%%%%%%%%% Figure 2 %%%%%%%%%%%%%%%%%%%%%%%%%%%%%%% 
\begin{figure}
\resizebox{\hsize}{!}{\includegraphics{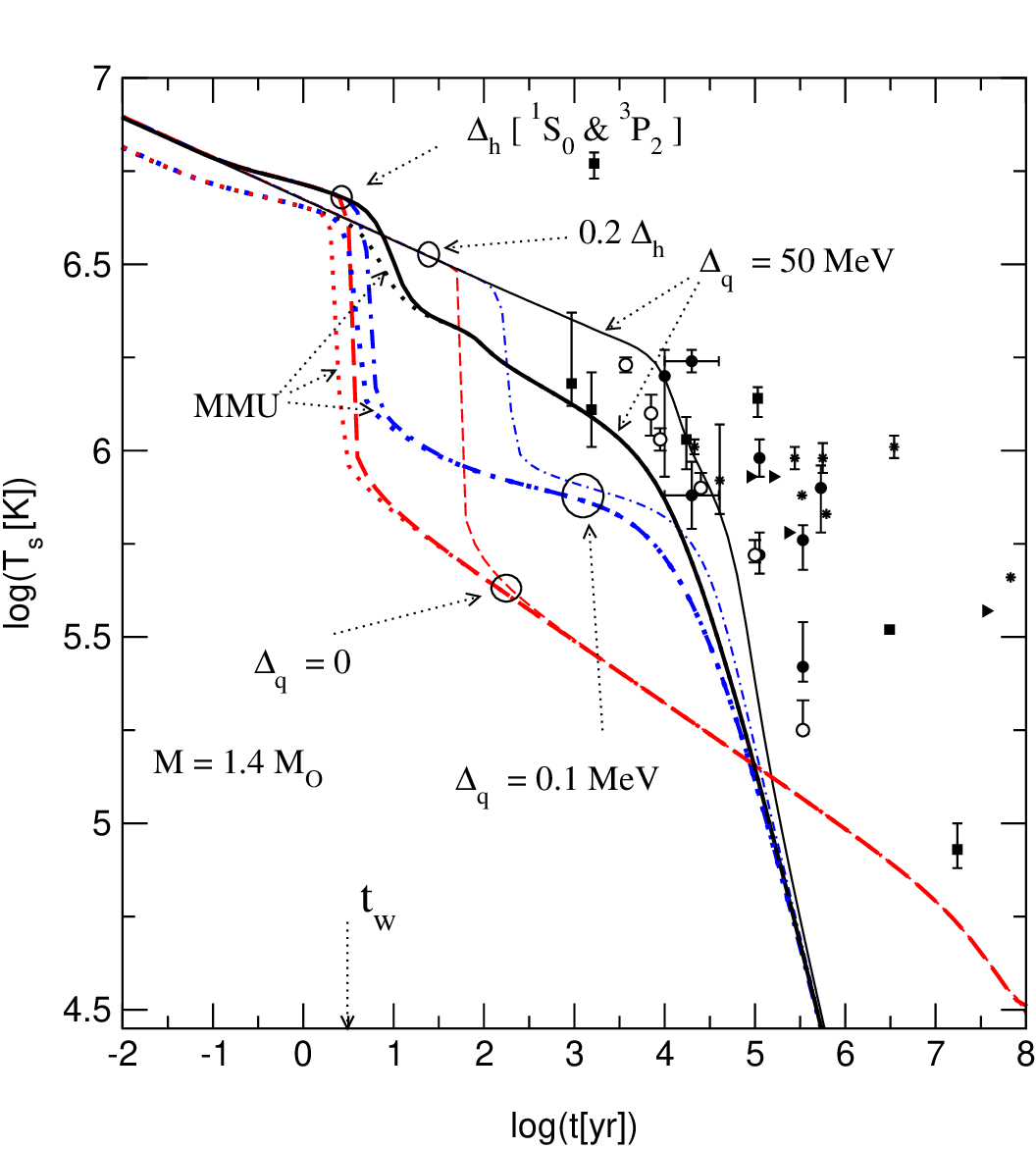}}
\caption{Evolution of the surface temperature $T_s$ of HNS 
with $M=1.4 M_{\odot}$ for $T_s\propto T_m^{2/3}$. Data points 
are from Schaab et al. (1999) (full symbols) and from Yakovlev et al. (2000)  
(empty symbols). 
$t_w$ is the typical time which is necessary for the cooling 
wave to pass through the crust.}
\label{2scequal}
\end{figure} 
%%%%%%%%%%%%%%%%%%%%%%%%%%%%%%%%%%%%%%%%%%%%%%%%%%%%%%%%%%%%%%%%%%%%%%%%% 
As can be seen from Fig. 2, these effects have an influence on the cooling  
evolution only for $\mbox{lg}(t[{\rm yr}]) \lsim 2$ since our specific 
model equation of state does not allow for large nucleon densities in the 
hadron phase for the $M = 1.4~M_\odot$ neutron star under discussion.  
The effect would be more pronounced for larger star masses, a softer  
equaton of state for hadron matter and smaller values of the gaps in the  
hadronic phase. 
 
The unique asymptotic behaviour at $\mbox{lg}(t[{\rm yr}]) \geq 5$ for all  
the curves corresponding to finite values of the quark and nucleon gaps is  
due to a competition between normal electron contribution to the specific 
heat and the photon emissivity from the surface since small exponentials  
switch off all the processes related to paired particles.  
This tail is very sensitive to the interpolation law $T_s =f(T_m)$ used to  
simplify the consideration of the crust.  
The curves coincide at large times due to the uniquely chosen relation  
$T_s \propto T_m^{2/3}$. 
 
The curves for $\Delta_q = 0.1$ MeV demonstrate an intermediate cooling 
behaviour between those for $\Delta_q=50$ MeV and $\Delta_q=0$. 
Heat transport becomes not efficient after first $5 \div 10$ yr.  
The subsequent $10^4$ yr evolution is 
governed by QDU processes and quark specific heat being only moderately 
suppressed by the gaps and by the rates of NPBF processes in the 
hadronic matter (up to $\mbox{lg}(t[{\rm yr}]) \leq 2.5$).  
At $\mbox{lg}(t[{\rm yr}]) \geq 4$ begins the photon cooling era. 
 
The curves for  normal quark matter ($\Delta_q =0$) are governed by 
the heat transport at times $t \lsim 5$ yr and then by QDU processes and the 
quark specific heat. The NPBF processes are important up to 
$\mbox{lg}(t[{\rm yr}])\leq 2$, the photon era is delayed up to  
$\mbox{lg}(t[{\rm yr}])\geq 7$.  
For times smaller than $t_w$ (see Fig. 2) the heat transport is delayed within 
the crust area (\cite{LPPH91}). Since we, for simplicity, disregarded this 
delay in our heat transport treatment, for such small times the curves should 
be interpreted as the $T_m(t)$ dependence scaled to guide the eye by the same 
law $\propto T_m^{2/3}$, as $T_s$.

\section{Temperature evolution of QCNS} 
\subsection{Temperature profiles $T(r,t)$} 
 
As can be seen from the lower panel of Fig. 3 for our large gap example  
($\Delta_q =50$MeV), where $Y_e\neq 0$ is determined from equation of 
state, the QCNS cools down rather smoothly from its surface during the first  
300 yr.  
%%%%%%%%%%%%%%%%%%%%%%%%%%%%%%%%% Figure 3 %%%%%%%%%%%%%%%%%%%%%%%%%%%%%%% 
\begin{figure}
\resizebox{\hsize}{!}{\includegraphics{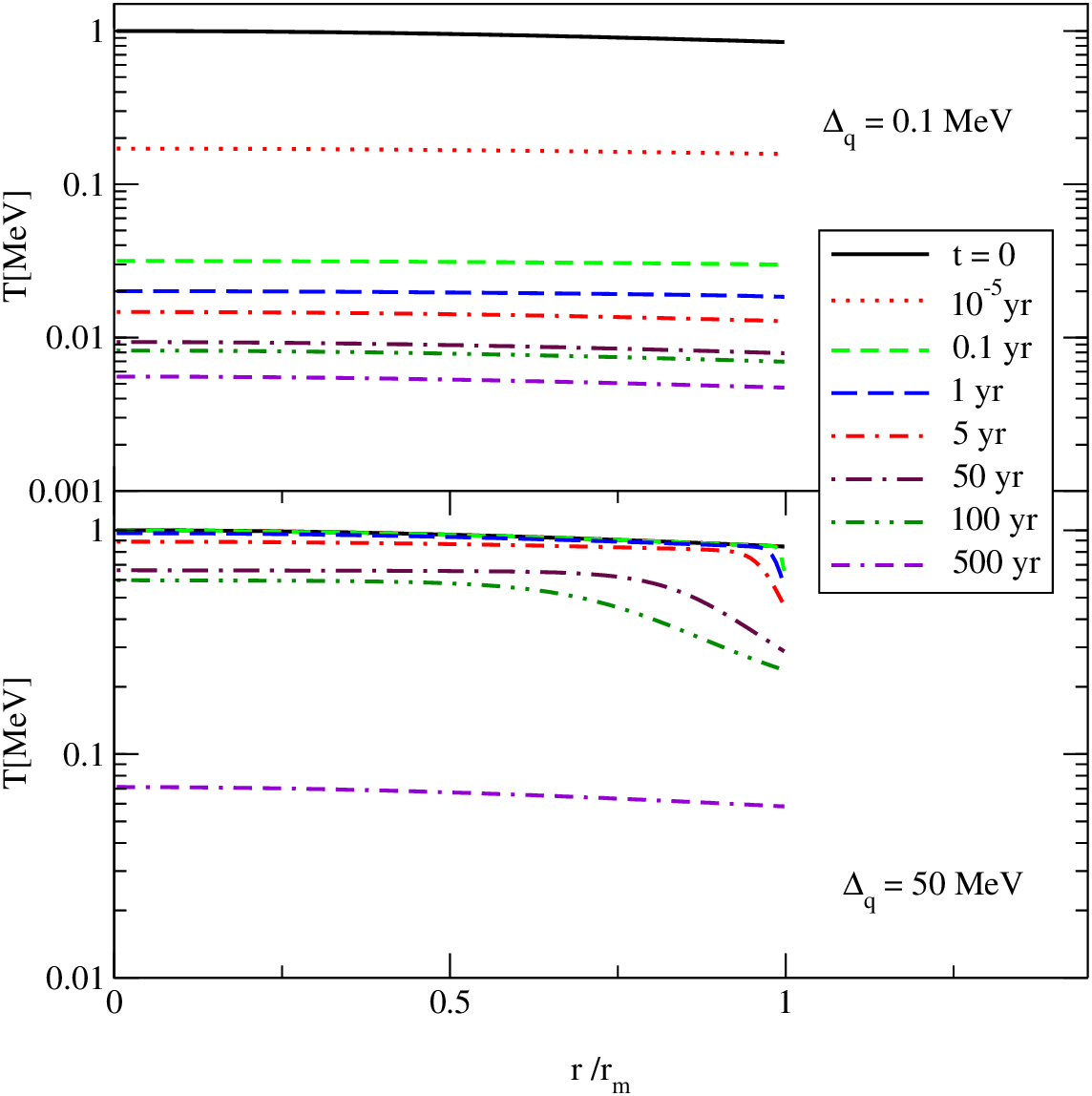}}
\caption{Early evolution of temperature profiles for a QCNS of  
$M=1.4 M_\odot$  with large (lower panel), and small 
(upper panel) diquark pairing gaps.}  
\label{Tprofq}
\end{figure} 
%%%%%%%%%%%%%%%%%%%%%%%%%%%%%%%%%%%%%%%%%%%%%%%%%%%%%%%%%%%%%%%%%%%%%%%%%% 
 
All the small gap examples exhibit about the same behaviour as the case  
$\Delta_q = 0.1$ MeV which we present in the upper panel of Fig. 3. 
Opposite to the large gap case, the transport processes are very efficient  
and a homogeneous temperature profile is recovered at $t\ll 10^{-5}$ yr.  
In reality in the latter case $T(r_m,t) \simeq T(r_m,0)$  
at times $\ll$ 1 yr with a drop of $T(r)$ in a narrow 
interval near $r/r_m =1$ and we disregard this peculiar behaviour 
as we have explained this above discussing HNS. 
 
%%%%%%%%%%%%%%%%%%%%%%%%%%%%%%%%% Figure 4  %%%%%%%%%%%%%%%%%%%%%%%%%%%%%%% 
\begin{figure}
\resizebox{\hsize}{!}{\includegraphics{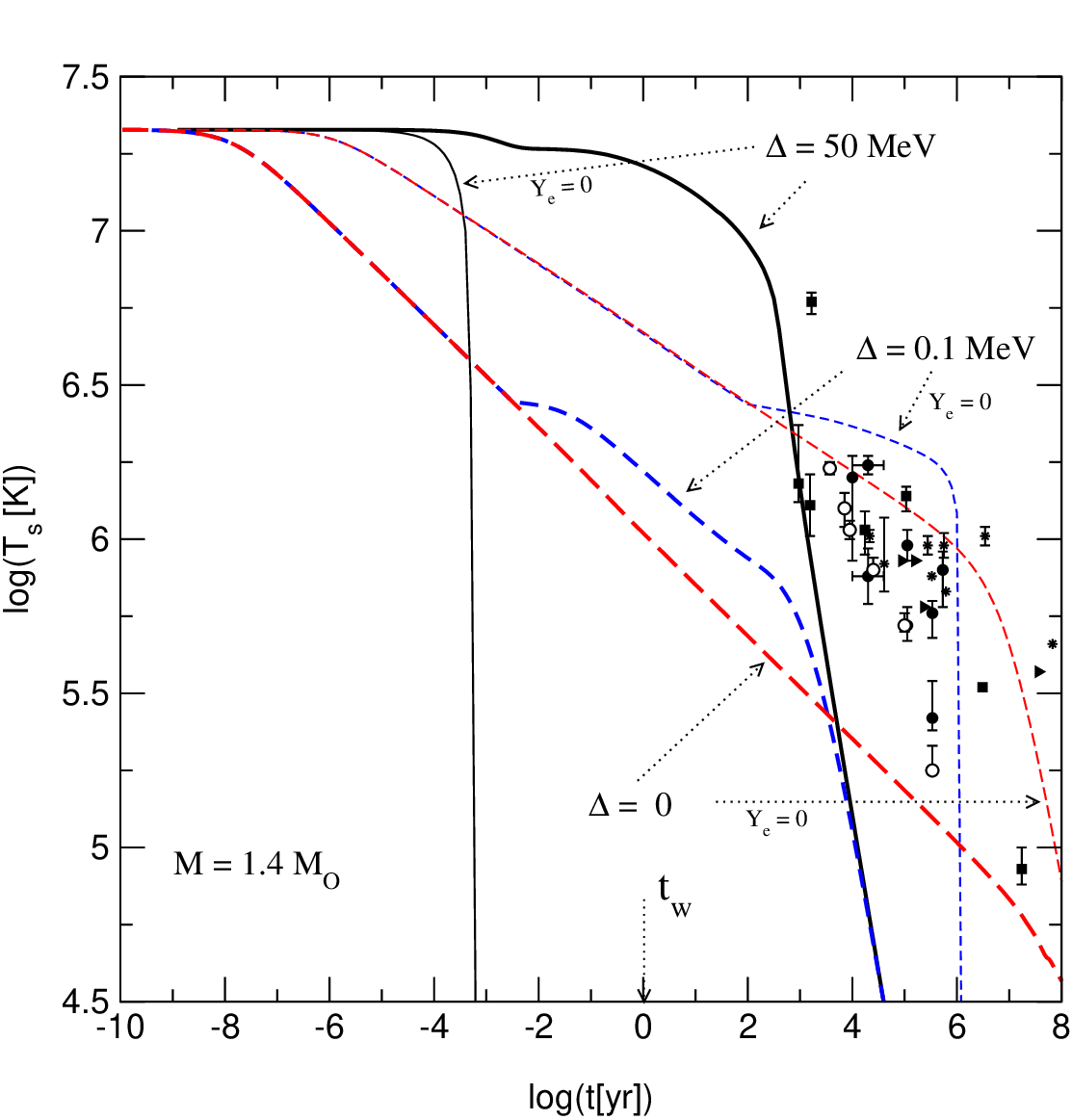}}
\caption{Evolution of the surface temperature $T_s$ for the QCNS 
with $M=1.4 M_{\odot}$ for $T_s\propto T_m^{2/3}$.} 
\label{2sc}
\end{figure} 
%%%%%%%%%%%%%%%%%%%%%%%%%%%%%%%%%%%%%%%%%%%%%%%%%%%%%%%%%%%%%%%%%%%%%%%%%% 
 
\subsection{Evolution of the surface temperature} 
 
In Fig.4 we demonstrate the evolution of the surface temperature  
$\mbox{lg}~T_s$ vs. $\mbox{lg}~t$ for representative cases. 
The curve for $\Delta_q =50$ MeV ($Y_e \neq 0$) shows a 300 yr delayed cooling 
which then is controlled by the electron specific heat and the photon 
emissivity from the surface, as it was stated in (\cite{bkv99}). 
However, in difference with the corresponding curve of (\cite{bkv99}) 
we have a much larger typical cooling time scale ($t \sim 300$ yr) 
since in that work the simplifying approximation of a homogeneous  
temperature profile has been used. 
For the low gap limit $\Delta_q \ll 10$ MeV in this particular QCNS case 
there is no heat transport delay and our results coincide with those 
obtained in (\cite{bkv99}).  
Unimportant differences are only due to the inhomogeneous density profile and 
a different value of $Y_e (n)$ which follows in our case  
from an actual calculation for the given equation of state. 
The curve corresponding to $\Delta_q =0.1$ MeV ($Y_e \neq 0$) 
from very early times is due to a competition between QDU emissivity and the 
quark specific heat, both being only moderately suppressed 
whereas late time asymptotics, as in the large gap example, relates to the 
photon era. A similar trend is present for normal quark matter 
staying below the corresponding curves for $\Delta_q =0.1$ MeV 
due to the absence of the $\zeta$ suppression factors 
in the former case. 
At $t> 10^{8}$ yr both cases have a different photon era asymptotic behaviour 
whereas all finite gap curves have the same large time asymptotics when  
$Y_e \neq 0$. 
 
At $Y_e =0$ the QDU processes are absent and the large time  
asymptotics are governed by the gluon-photon specific heat which is a 
nonlinear function of the temperature and the photon emissivity.  
Due to this nonlinearity the large time asymptotics are 
different in all the cases under consideration. 
All the results for $Y_e =0$ (including case of large gaps) coincide with  
those obtained in (\cite{bkv99}) although the sharp fall of the curve for the  
large gap case is in reality controlled by the heat conductivity of the crust  
that will lead to a cooling delay within the first year of evolution 
(\cite{Pizzochero 1991}) as in Fig. 2 the curves for $t<t_w$ should be 
interpreted as $T_m(t)$ dependence scaled by the Tsuruta law.
 
\section{Conclusion}   
 
For the CFL phase with large quark gap, which was expected to exhibit  
the most prominent manifestations of colour superconductivity in HNS and  
QCNS, we have found an essential delay of the cooling during the first  
$50 \div 300$ yr due to a dramatic suppression of the heat conductivity in  
the quark matter region.  
This delay makes the cooling of QCNS not as rapid as 
one could expect when ignoring the heat transport. 
In HNS compared to QCNS (large gaps) an additional delay of the subsequent  
cooling evolution comes from the processes in pure hadronic matter. 
In spite of that we found still too fast cooling for those objects. 
Therefore, with the CFL phase of large quark gap it seems   
rather difficult to explain the majority of the presently known 
data both in the cases of the HNS and QCNS, whereas in the case 
of pure hadronic stars the available data are much better fitted within the  
same model for the hadronic matter we used here. 
For 2SC (3SC) phases we expect analogous behaviour to that demonstrated 
by $\Delta_q =0$ since QDU processes on unpaired quarks are then allowed.    
We however do not exclude that new observations may lead to 
lower surface temperatures for some supernova remnants and will be 
better consistent with the model which also needs further improvements. 
On the other hand, if future observations will show very large temperatures  
for young compact stars they could be interpreted as a manifestation of  
large gap color superconductivity in the interiors of these objects. 
  
\begin{acknowledgements} 
Research of H.G. and D.B. was supported in part by the Volkswagen Stiftung   
under grant no.\ I/71 226 and by DFG under grant no. 436 ARM 17/1/00. 
H.G. acknowledges the hospitality of the Department of Physics at the  
University of Rostock where this research has been performed. D.N.V.  
is grateful for hospitality and support of GSI Darmstadt.  
D.B. acknowledges 
support of the Department of Energy for his participation in the INT program  
on ``QCD at Nonzero Baryon Density'' (INT-00-1) at the University of 
Washington as well as a fellowship at the ECT$^*$ Trento where this work was  
completed.   
We thank Th. Kl\"ahn for his contributions during early stages of this work; 
M. Colpi, M. Prakash, K. Rajagopal, S. Reddy, A. Sedrakian and F. Weber  
for their discussions during the Workshop ``Physics of Neutron Star 
Interiors'' at the ECT$^*$.  
\end{acknowledgements}

\end{document}